\documentclass[aps,prl,twocolumn,groupedaddress]{revtex4}
\usepackage[pdftex]{graphicx,epsfig}

\begin{document}

\title{Self-organized global control of carbon emissions}

\author
{Zhenyuan Zhao$^{1}$, Dan Fenn$^{2}$, Pak Ming Hui$^{3}$ and Neil
F. Johnson$^{1}$}
\affiliation{$^{1}$Physics Department, University of Miami, Coral Gables, FL 33126, U.S.A.\\
$^{2}$Oxford Centre for Industrial and Applied Maths, Oxford University, Oxford OX1 3PU, U.K.\\
$^{3}$Department of Physics, The Chinese University of Hong Kong, Shatin, Hong Kong}

\date{\today}

\begin{abstract}
There is much disagreement concerning how best to control global carbon emissions. We explore quantitatively how different control schemes affect the collective emission dynamics of a population of emitting entities. We uncover a complex trade-off which arises between average emissions (affecting the
global climate), peak pollution levels (affecting citizens'
everyday health), industrial efficiency (affecting the nation's
economy), frequency of institutional intervention (affecting
governmental costs), common information (affecting trading
behavior) and market volatility (affecting financial stability). Our findings predict that a self-organized free-market approach at the level of a sector,
state, country or continent, can provide better control than a
top-down regulated scheme in terms of market volatility and
monthly pollution peaks.
\end{abstract}

\maketitle

A CO$_2$ emissions level of $\leq 500$ ppm \cite{SternReview} would set the probability of
a potentially catastrophic 5$^\circ$C warming at
3$\%$\cite{SternReview}. At a
recent G8 summit, leaders agreed to `strongly consider' at least
halving global emissions by 2050\cite{bbc}. However, there is still no national or international consensus on how these
reductions can be systematically achieved and
maintained\cite{emissionstrading}, nor is there any deep quantitative understanding of the trade-offs which could arise at the local and global level.
Given the recent instabilities in global financial
markets and apparent inevitability of human
irrationality\cite{Greenspan}, it is also unclear whether a free-market approach 
can ever be trusted\cite{irrational}.

Here we analyze a simple, yet realistic dynamical model of a
competitive emissions market which allows us to investigate the
simultaneous interplay between myriad competing real-world
factors. Our model is a non-trivial generalization of the  El Farol bar
problem\cite{arthur} which has attracted much attention among physicists\cite{challet,us,neil}.
In addition to offering the physics community a novel generalization and application of the El Farol model, we believe that our work provides the first unified, quantitative discussion of the
underlying trade-offs between average emissions,
instantaneous peak pollution levels, market stability, efficiency
of production, and common information. Our model  predicts
that a completely self-organized emissions market with collective
competition and no top-down management, can offer distinct
advantages over a managed system in terms of peak emission values.
Although helpful with respect to the mean monthly emission,
top-down monthly management can by contrast induce a far bigger
volatility and hence aggravate the uncertainty in emissions. 

\begin{figure}
\center
\includegraphics[width=0.49\textwidth]{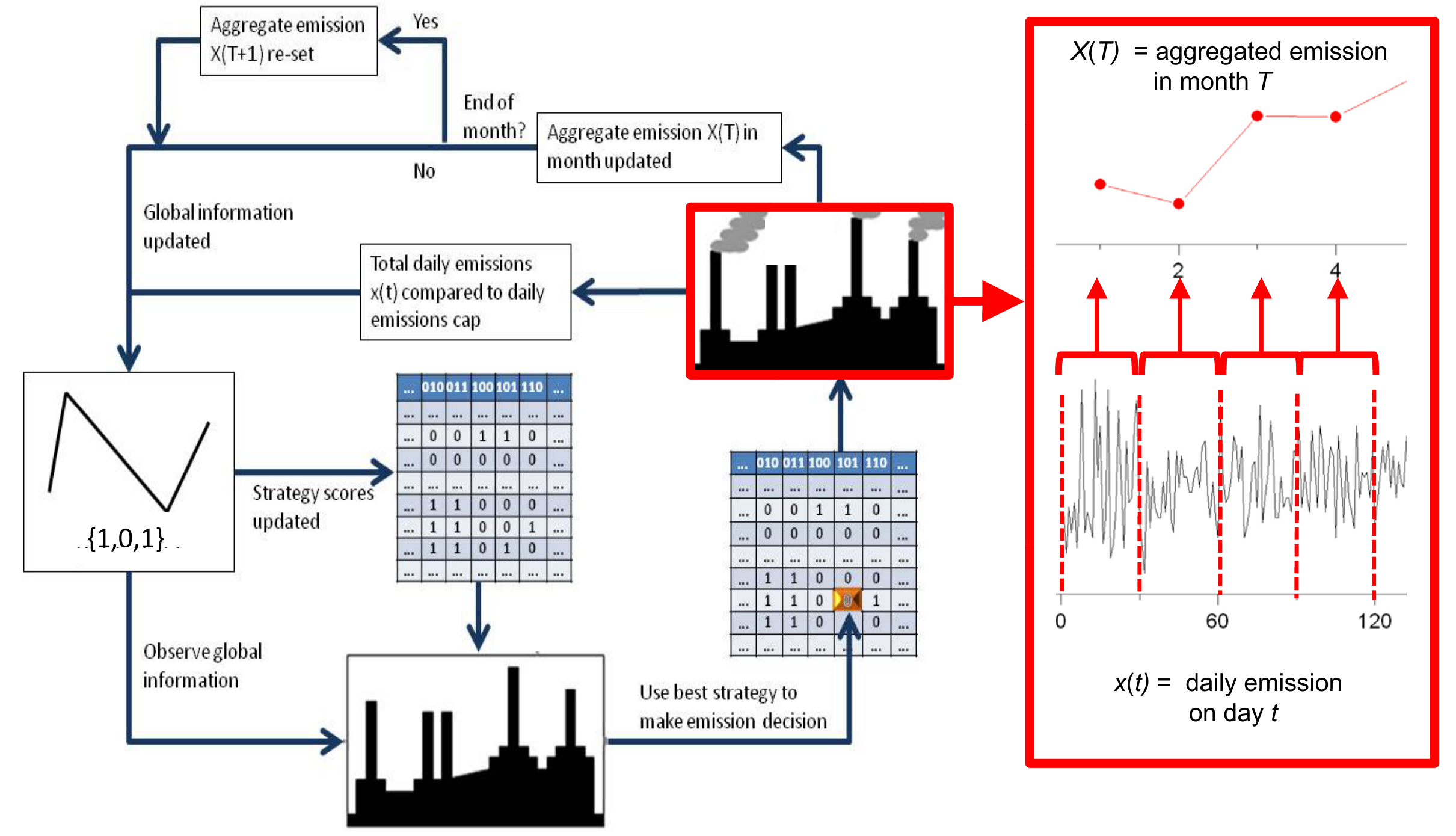}
\caption{(Color) Schematic diagram of the carbon market model and the
resulting daily time series $x(t)$ for day $t=1,2,\dots$, together
with the aggregated monthly time-series $X(T)$ for month
$T=1,2,\dots$.}
\end{figure}

Figure 1 shows a schematic describing our generic emissions
scenario, comprising $N$ emitters (e.g. companies) who each decide
whether to emit or not during a particular timestep $t$ (e.g.
day). All companies are assumed to have the same emission
capabilities (i.e., one unit of carbon each timestep). The
system's (e.g. national) safe emission level is $L$ over some
period $\Delta t$ (e.g. month $\Delta t =30$), with successive periods (e.g. months) labelled $T=1,2,\dots$. Hence the average emission cap per
timestep (e.g. day) is $\bar{L}=L/\Delta t$. Depending on the top-down management
infrastructure of interest, the emitters could equally well be
industries within a sector, companies within a state, states
within a country, countries within a continent, or countries or
continents within some global organization -- likewise, the
relevant timescales $t$ and $T$ need not be days and months
respectively. From a governmental perspective, the ideal outcome
would be that the total emission each month $X(T)$ is exactly
equal to $L$ units of carbon pollutants: If $X(T)>L$ then too much
carbon dioxide is emitted into the atmosphere, while $X(T)<L$
means that the nation has wasted some of its allowed production
capacity\cite{emissionstrading}. Companies are rewarded in some generic
way (e.g. favorable public opinion, or a monetary compensation)
for choosing to emit on low-pollution days ($x(t')\leq \bar{L}$)
or abstaining from emitting on high-pollution days ($x(t')>
\bar{L}$), and receive punishments otherwise. Each day's outcome
is represented in terms of its collective emission: 1 if
$x(t')\leq \bar{L}$ for a given $t'$ and 0 if $x(t')>\bar{L}$.
Companies rely on common, publicly disclosed information when
deciding whether or not to emit at a given timestep. We take this
common information to be dominated by the previous $m$ days'
outcomes, a bit-string of length $m$ compromised by 0 or 1, but in
principle it could include other information from government,
public or other competitors. The fact that all participants have
access to, and use, the same information, can generate
correlations between their actions. A strategy is a specific
prediction 0 or 1 (and hence action, emit or don't emit) for each
of the ${2^m}$ possible information bit-strings, hence there are
$2^{2^m}$ strategies. Companies randomly select $s$ strategies
from the strategy space with repetitions allowed during the
assignment. Each company uses its best performing strategy at a
given timestep, with an individual strategy's score updated by
$+1$ ($-1$) at a given timestep, if it would have made the correct
(incorrect) decision. The correct decisions are emitting (not
emitting) when the cap is not exceeded (exceeded), and vice versa
for incorrect decisions. Tied best-performing strategies are
broken by random choices. Our setup therefore incorporates the
generic complex system features of Arthur's El Farol problem and
Challet and Zhang's binary
version\cite{arthur,challet,us,neil}.
Most importantly, companies do not communicate directly among
themselves, nor do they need to know the number of competitors
around, nor are they managed by some governmental entity. Instead,
by competing to emit, they interact through the common information
that their collective actions create. There is recent independent
evidence that groups of human do indeed employ such general
decision-based mechanisms as in Fig. 1\cite{PNASmg}. More
generally, our model mimics a simple cap-and-trade scenario in
which emitters who decide to emit on a given day immediately
purchase a permit to do so. The less emitters per day, the lower
the demand for permits, and hence the lower that day's permit
price, and vice versa. Hence the time-series of emissions $x(t)$
mimics the time-series of permit prices.

\begin{figure}
\center\includegraphics[width=0.49\textwidth]{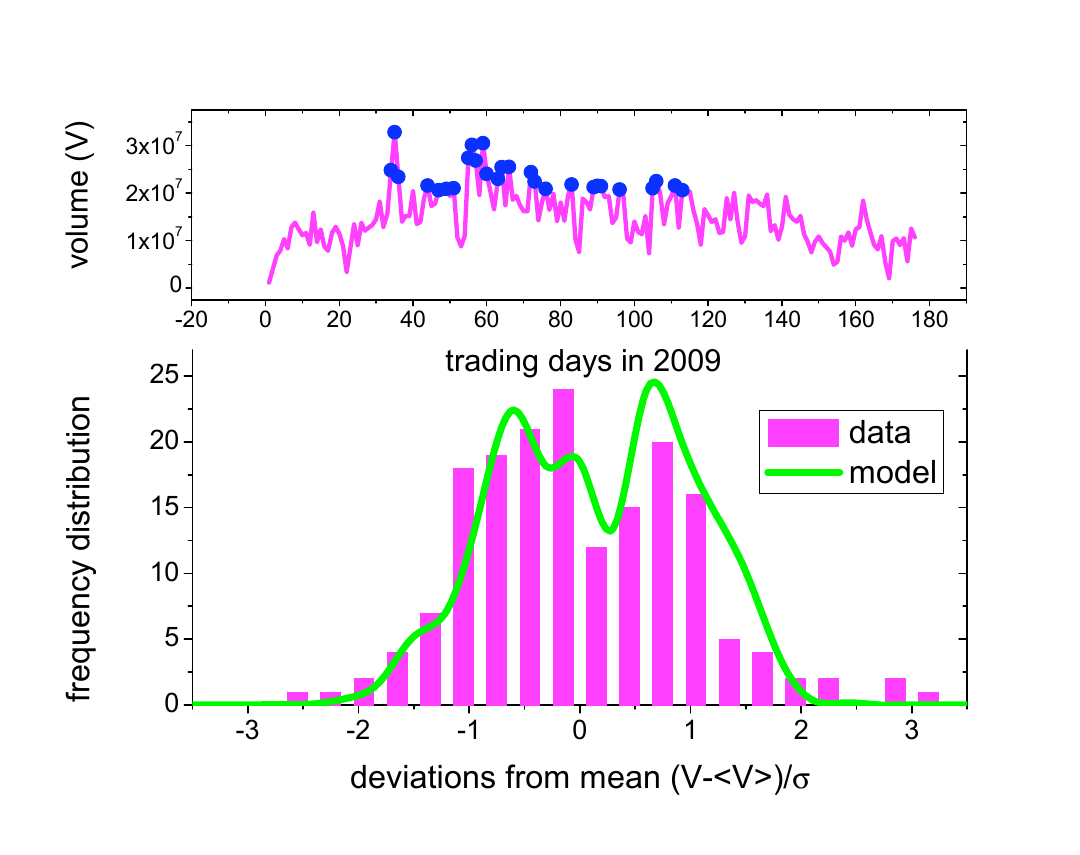}
\caption{(Color) The upper panel is the daily time series of the first
$176$ trading days' total volume in $2009$ from ECX EUA futures
contract, in tonnes of CO$_2$ (EU allowances). The lower panel
shows the empirical frequency distribution of $(V-\langle V
\rangle)/\sigma)$, as compared to our model $m = 4$, averaged over
$100$ runs. Our model is a good representation of the empirical
data, but has the added advantage that it eliminates the extreme
peaks observed in the EU market (indicated by circles in the top
panel). This suggests that our unmanaged free market mechanism
would provide tighter control than the existing EU market.}
\end{figure}

Figure 2 shows that this simple model is capable of reproducing
the highly irregular, non-Gaussian distribution of the 2009 EU
carbon market to date. The model parameters and their values have
a direct and reasonable interpretation: $N=100$ suggests that the
actions of approximately one hundred entities (e.g. large
companies) is moving the market in 2009, and hence visibly
impacting the overall emissions; $m=4$ suggests that just less
than one week of prior outcomes is considered relevant for making
a decision; $s=6$ suggests that individual entities are using
approximately six strategies to make a decision about whether to
buy a permit and hence emit. Choosing to emit is
equivalent to buying a permit and using it on that day -- if less
people apply on a given day, the permit price is low  which means that
the time-series of the number of emitters and the price mimic each
other. Hence as a surrogate of the actual daily emissions, we have taken
the daily carbon price to represent the daily demand for
permission to emit, and hence the resulting volume of emissions.
The quantity displayed, $(V-\langle
V\rangle)/\sigma$, is independent of the number of participants
$N$, for large $N$. We note that our model shows a smaller
occurrence of extreme events than the empirical data, suggesting
that our competitive, self-organized setup might provide better
control of large fluctuations than the present EU scheme which is
operating. If the distribution were more Gaussian-like as in
regular financial markets, this would suggest that the market
should contain many noisy speculators -- however, the multi-modal
form in Fig. 2 implies that this is not the case.

\begin{figure}
\center\includegraphics[width=0.49\textwidth]{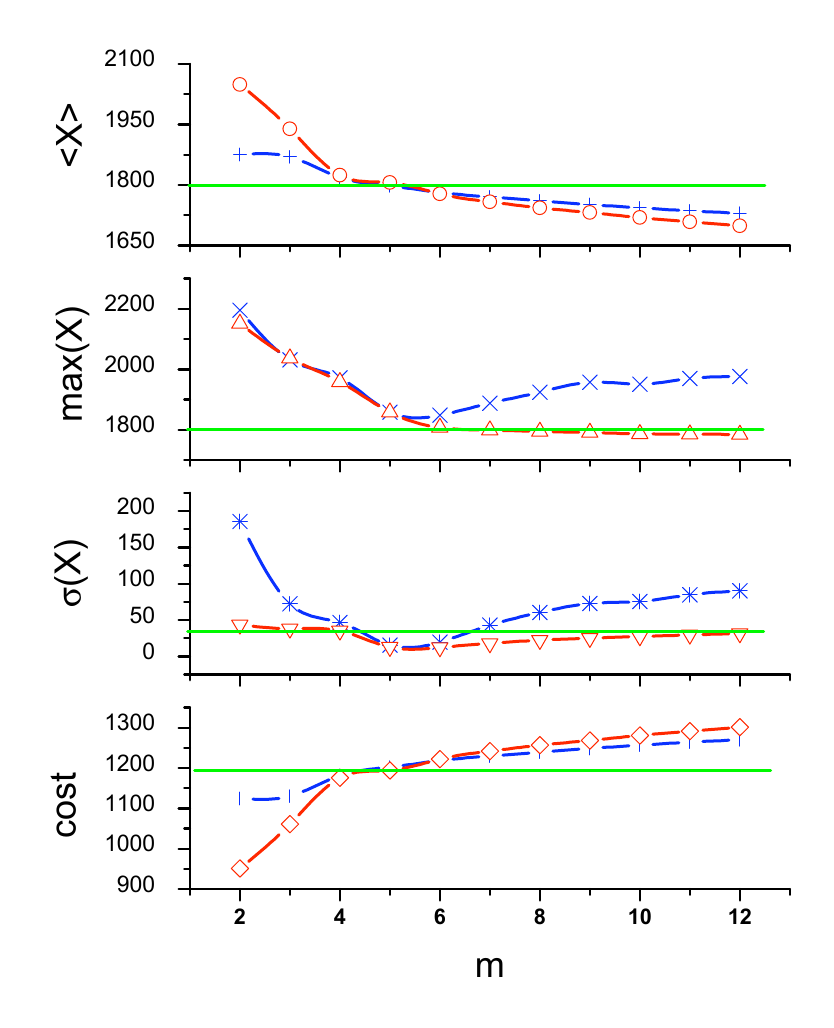}
\caption{(Color) Monthly emissions for our model, for $N=100$, $s=6$. Top
to bottom: the mean monthly emission $\langle X \rangle$, the
maximum (i.e. peak) monthly emission, the monthly volatility
$\sigma(X)$, and the governmental cost of compensation to
companies for not emitting. Red: unmanaged system. Blue: managed
system. Green: random result for learning $p=\bar{L}/N$.}
\end{figure}

Figure 3 compares the predictions of our model for monthly
emissions between an unmanaged (red curve) and managed (blue
curve) system, as a function of the amount of common information
about previous outcomes (i.e. $m$). The average
daily emissions cap is $\bar{L}=60$. In the managed system, at the
end of month $T$, the government will reduce or increase the
emissions capacity ${L}(T+1)$ for month $T+1$ by the amount that
the aggregated emissions $X(T)$ was above or below ${L}(T)$. In
the unmanaged system, there is no such external control and hence
${L}$ is constant. The overall system performance can be assessed
through the time-series for monthly emissions $X(T)$ (Fig. 1): In
particular (top to bottom in Fig. 3) the mean $\langle X\rangle$,
the maximum ${\rm max}(X)$ (where ${\rm max}(X)$ is the largest
monthly emission value during the time-window of the numerical
simulation) over some fixed period (e.g. a year), and the standard
deviation (i.e. volatility $\sigma(X)$) about the mean. An
interesting comparison system is obtained by considering the
`random' case of an unmanaged system in which companies decide to
emit by tossing a coin each day. In the absence of any learning
(i.e. the system is non-adaptive) every decision is an independent
coin toss and hence the a priori probability to emit would be
$p=0.5$. However if the entities are gradually able to learn from
the feedback of the previous experience and adapt to the ideal
ratio $\bar{L}/N$ (at least, at the collective level) then
$p=\bar{L}/N$ yielding the green curves shown in Fig. 3.

The mean monthly emission $\langle X\rangle$ decreases
monotonically as $m$ increases for both systems. The monthly
control exerted in the managed system pulls the value closer to
the capacity limit of $1800$ than for the unmanaged system.
However, this improved performance due to top-down management is
accompanied by a significantly higher volatility for $m>6$ as well
as a significantly higher peak pollution level. Indeed, the
managed system does {\em worse} than both the unmanaged system
{\em and} the random system with learning. This is because the
month-by-month adjustment to $\bar L$ induces a delayed
oscillatory effect which in turn generates significant volatility.
A transition occurs around  $m \sim 4$ where all the curves seem
to cross the green (i.e. random learning) curve. This $m$ value
coincides with the system's dynamical de Bruijn path (which has
duration $2.2^m$ \cite{m4}) becoming equal to the finite duration
of the emission interval (i.e. 30 days, hence $2.2^m\sim 30$ which
yields $m \sim 4$). This precedes a minimum in the volatility
around $m\sim 5$ for both managed {\em and} unmanaged systems,
which is smaller than for random learning. As for the El Farol
problem\cite{arthur,challet,us,neil},
this unintentional collective cooperation emerges as a result of
cancellation between the actions of crowds of emitters using one
strategy, and anticrowds using the exact opposite strategy. The
cost result (bottom panel) reflects a simple one-unit payout given
to any company not emitting on a given day. The choice of emitting
or not-emitting becomes essentially cost-neutral to a given
company -- however for public relations reasons, and because they
want to stay active in business, each company still continues to
compete. A higher $\langle X\rangle$ hence incurs a lower cost.

\begin{figure}
\center\includegraphics[width=0.48\textwidth]{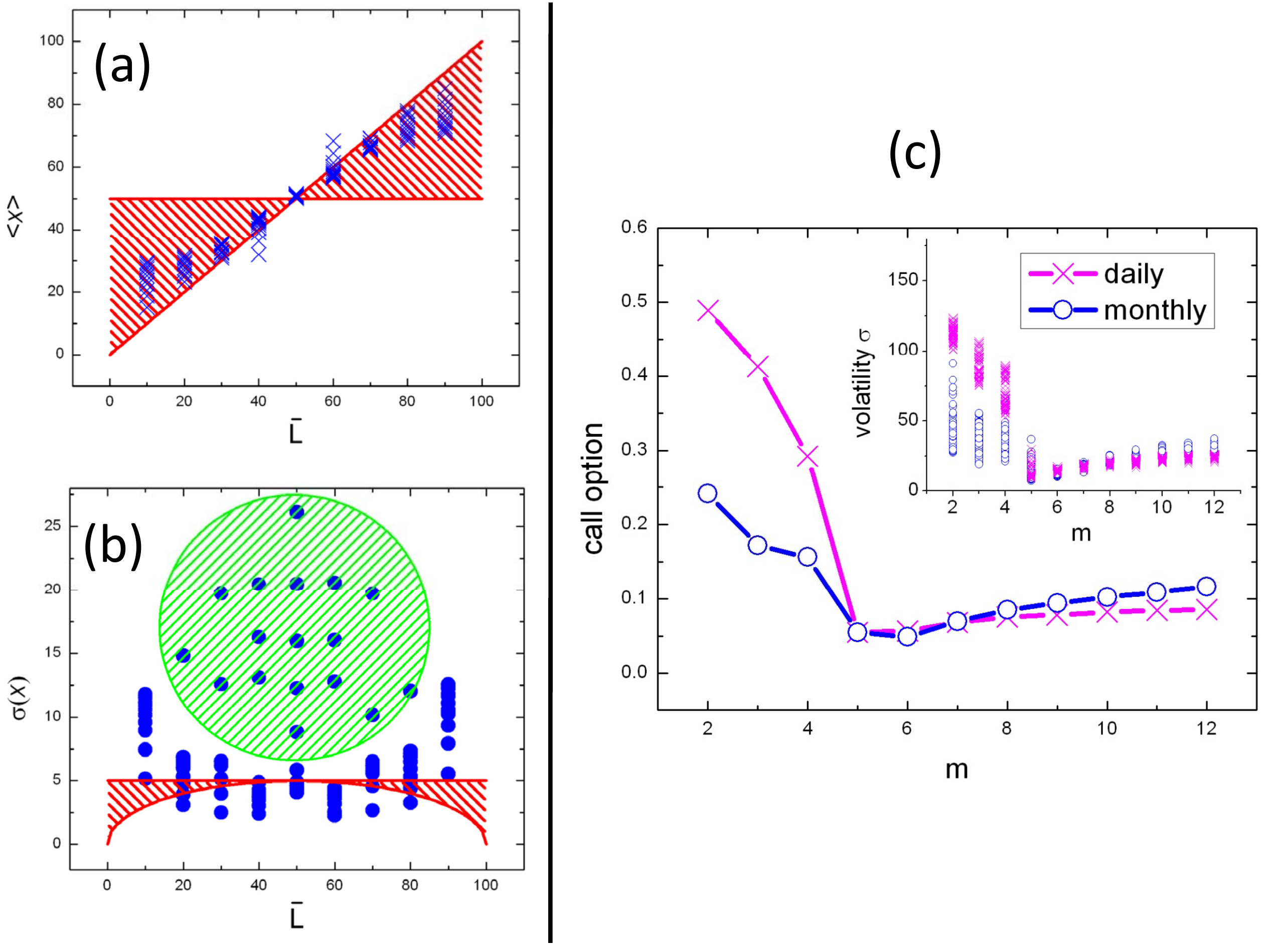}
\caption{(Color) Daily emissions for our model with $m$
varying from 2 to 12 ($s=6$). Red shaded region is the analytically obtained zone of learning
(see text). (a) Mean emission. Horizontal boundary line
corresponds to no learning, diagonal boundary line (slope of
unity) is for learning. (b) Volatility. Horizontal boundary line
corresponds to no learning, convex curve is for learning.
Green shaded circle shows low $m$, high volatility region due to
crowding in strategy space. (c) European call option prices for
different measurements of volatility according to the standard
derivative pricing theory (i.e. Black-Scholes equation). Risk-free
interest rate $r=0$, current value $x$ is set to the individual
mean divided by 100, and strike price $X_s$ equals $\bar{L}/100$.
The volatility $\sigma$ is scaled by $1/\sqrt{100}$. The time is
one day before expiration. Blue curve uses the results from
monthly measurement, while purple one is from daily measurement.
The insert shows the anomalous scaling of volatility of emissions,
from daily to monthly scales. There are $64$ runs for each $m$.
The purple crosses are the standard deviation calculated by taking
the daily volatility and multiplying by $\sqrt{30}$, which would
be {\em exactly} equal to the monthly volatility if the
time-series followed a random walk.}
\end{figure}

Figure 4 shows the model's daily emission (Fig. 4(a), crosses) and
volatility (Fig. 4(b), crosses) as a function of the daily
emissions cap $\bar L$. The red shaded area in Figs. 4(a) and (b)
is the `learning zone' bounded by the two analytically obtained limits of no learning
($p=0.5$, the probability that a company is going to emit at a
timestep, horizontal red line) and learning ($p=\bar{L}/N$, red
diagonal line in Fig. 4(a) and convex curve in Fig. 4(b)). The
standard deviation for daily emissions in the random case, is
given analytically by the usual binomial form, i.e. $[Np(1-p)]^{\frac{1}{2}}$. Using the
lower bound value $p=\bar{L}/N$ yields the convex curve
$[\bar{L}(1-\bar{L}/N)]^{\frac{1}{2}}$, while using the upper
bound value $p=0.5$ yields the horizontal line $0.5
N^{\frac{1}{2}}=5$. The model's mean emission values (crosses) lie
within the shaded area in Fig. 4(a) but are closest to the limit
$p=\bar{L}/N$, thereby demonstrating that the unmanaged,
self-organized market collectively learns. For intermediate $\bar
L$ values (Fig. 4(b)) the corresponding volatility tends to be
smaller than the random value, however it moves above it for very
large or small $\bar L$. For small $m$ values, which corresponds
to the crowded regime of the strategy space, numerical runs can
show significantly large volatilities (green circle).

Figure 4(c) explores the implications of our results for the {\em
derivative} emissions markets. If emission markets follow the path
of the mature non-emission financial markets, it is likely that
such derivatives (e.g. options) markets will become as large, or
even larger, than the primary emissions market itself\cite{neil}. In this
respect, our findings serve as a warning of the dangers of simply
applying standard financial theory for such derivative
instruments\cite{neil}. Standard option pricing theory uses the
volatility over a given time increment as the input to the
Black-Scholes pricing formula\cite{price}. This assumes that the
market approximates to a random walk and hence that the monthly
volatility over $\Delta t$ timesteps is simply the volatility over
one timestep multiplied by $\sqrt {\Delta t}$. Figure 4(c) shows
not only that this is incorrect (see inset), but also that the
discrepancy depends on the amount of common information $m$ -- and
that as a consequence, the price of a call option (Fig. 4(c)) can
be mispriced according to whether daily or monthly volatility
estimates are used. This opens up an intriguing but dangerous
situation in the event of any abnormal periods in the market:
Following some external news event (e.g. collapse of an oil
company), it may happen that the number of previous days' outcomes
that are thought relevant, becomes very small (i.e. $m\rightarrow
0$). As shown, the corresponding mispricing then becomes huge,
leading to possible financial instabilities.

Despite recent skepticism surrounding the stability of free markets, our analysis predicts that an {\em unmanaged}
carbon emissions market can provide significant advantages over a managed one.  For a given sector, state, country or
continent, our model helps identify the appropriate
degree of governmental management such that annual {\em global}
emissions targets are achieved, while simultaneously allowing for
individual choice regarding the trade-off between {\em local}
social issues as listed in the abstract. Finally, we have checked that our main conclusions are reasonably robust to different sets of parameter values.


\begin{thebibliography}{99}
\bibitem{SternReview} N. Stern, {\em The Economics of Climate Change: The Stern Review}. (Cambridge Univ. Press, Cambridge,
2007); {\em A Blueprint for a Safer Planet}. (Random House, London,
2009), pp. 39; A.D. Ellerman, {\em et al.} {\em Markets for Clean
Air: The U.S. Acid Rain Program}. (Cambridge Univ. Press,
Cambridge, 2000).
\bibitem{bbc} See news.bbc.co.uk/1/hi/world/europe/6732787.stm
\bibitem{emissionstrading} {\em Emission Trading: Environmental Policy's New
Approach}, Eds. R.F. Kosobud, D.L. Schreder,
and H.M. Biggs (John Wiley $\&$ Sons, Inc., New York, 2000); {\em Voluntary Carbon Markets: An International Business Guide
to What They Are and How They Work} Eds. R. Bayon, A. Hawn, and K. Hamilton (Earthscan Publications Ltd.,
London, 2009); {\em Emission Trading: institutional Design,
Decision Making and Corporate Strategies}, Eds. R. Antes, B. Hansjurgens,
and P. Letmathe (Springer, New York,
2008).
\bibitem{Greenspan} See news.bbc.co.uk/2/hi/business/8244600.stm
\bibitem{irrational} J. O'Brien, {\em Engineering a Financial Bloodbath}. (World Scientific, Singapore, 2009).
\bibitem{arthur} W.B. Arthur, {\em Amer. Econ. Assoc. Papers. Proc.} {\bf 84}, 405
(1994); Science {\bf 284}, pp. 107-109 (1999).
\bibitem{challet} D. Challet, and Y.C. Zhang, {\em Physica A} {\bf 246}, 407 (1997); D. Challet, M. Marsili, and Y.C. Zhang, {\em Minority
Games}. (Oxford Univ. Press, 2005); A.C.C. Collen, {\em The mathematical theory of Minority
Games}. (Oxford University Press, 2005); T.Galla, D. Sherrington, {\em J. Stat. Mech.} P 10009
(2005); D. Sherrington, E. Moro, and J.P. Garrahan, {\em Physica A} {\bf
311}, 527 (2002); T. Galla, and A. De Martino, {\em J. Phys. A: Math. and Theor.} {\bf 41}
324003 (2008).
\bibitem{us} N.F. Johnson, {\em et al.}  {\em Physica A} {\bf 258}, 230 (1998).
\bibitem{neil} N.F. Johnson, P. Jefferies, and P.M. Hui, {\em Financial Market Complexity}. (Oxford
Univ. Press, 2003).
\bibitem{PNASmg} W. Wang, Y. Chen, and J. Huang, {\em Proc. Natl. Acad. Sciences. U.S.A.} {\bf 106}, 8423 (2009).
\bibitem{m4} P. Jefferies, M.L. Hart, and N.F. Johnson, {\em Phys. Rev. E} {\bf 65}, 016105 (2001).
\bibitem{price} Price of the European
call option, using Black-Scholes equation, is\cite{neil}:
$V(x,t)=x\Phi[d_1]-X_se^{-r(t_0-t)}\Phi[d_2]$, where
$\Phi[z]=\frac{1}{\sqrt{2\pi}}\int^{z}_{-\infty}e^{-0.5y^2}$,
$d_1=\frac{ln(x/X_s)+(r+0.5\sigma^2)(t_0-t)}{\sigma\sqrt{t_0-t}}$,
$d_1=\frac{ln(x/X_s)+(r-0.5\sigma^2)(t_0-t)}{\sigma\sqrt{t_0-t}}$,
where $\sigma$ is the volatility.





\end{thebibliography}
\end{document}